# mSigSDK - private, at scale, computation of mutation signatures

Aaron Ge*, Tongwu Zhang, Yasmmin Côrtes Martins, Maria Teresa Landi, Brian Park, Kailing Chen, Jeya Balasubramanian, Jonas S Almeida

* aaron.ge@nih.gov

## Abstract

In our previous work, we demonstrated that it is feasible to perform analysis on mutation signature data without the need for downloads or installations and analyze individual patient data at scale without compromising privacy. Building on this foundation, we developed an in-browser Software Development Kit (a JavaScript SDK), mSigSDK, to facilitate the orchestration of distributed data processing workflows and graphic visualization of mutational signature analysis results. We strictly adhered to modern web computing standards, particularly the modularization standards set by the ECMAScript ES6 framework (JavaScript modules).

Our approach allows for the computation to be entirely performed by secure delegation to the computational resources of the user's own machine (in-browser), without any downloads or installations. The mSigSDK was developed primarily as a companion library to the mSig Portal resource of the National Cancer Institute Division of Cancer Epidemiology and Genetics (NIH/NCI/DCEG), with a focus on FAIR extensibility as components of other researchers' own data science constructs. Anticipated extensions include the programmatic operation of other mutation signature API ecosystems such as SIGNAL and COSMIC, advancing towards a data commons for mutational signature research (Grossman et al., 2016).

**Availability:**

Main Website: https://episphere.github.io/msig

Github: https://github.com/episphere/msig

Observable Notebooks: https://observablehq.com/@aaronge-2020/signature-extraction

mSigPortal (dependency): https://analysistools-dev.cancer.gov/mutational-signatures/

# Introduction

Mutational signatures are patterns of genetic mutations resulting from specific biological processes that can be identified by analyzing the DNA sequences and comparing them to reference genomes (Alexandrov et al., 2020). This information can then be used to identify underlying biological mechanisms driving cancer development, develop personalized treatment plans, and identify potential environmental factors or genetic predispositions contributing to cancer development. In addition to informing opportunities for precision prevention (Landi et al., 2021), mutational signatures can provide insight into the efficacy of cancer treatments and the potential for drug resistance (Pich et al., 2019).

To facilitate mutational signature research, Zhang et al. started developing a web-based platform at NCI called mSigPortal in 2023 (Zhang et al., 2023). mSigPortal has just been made public at https://analysistools.cancer.gov/mutational-signatures/#/, allowing users to explore curated mutational signature databases, visualize and compare reference signatures, perform mutational signature de novo extraction and decomposition analysis using state-of-the-art algorithms, systematically explore mutational signature activities and performance, and statistically analyze associations between mutational signature features and sample-level variables. The platform also offers access to the mutational signature-related data programmatically, through a HTTP REST Application Programming Interface (API), upon which the mSig SDK (Software Development Kit) reported here was developed (Zhang et al., 2023).

The reusability of data-intensive computational constructs, including assumptions that traditional (non-reactive) notebooks address it effectively, has become a major concern (Perkel, 2021). We argue here that the modularization of libraries operating data-intensive resource APIs offers a necessary foundation, and leads to far more reusable interactive analytics tools. In 2016, Wilkinson et al. introduced the Modern FAIR principles for stewardship of scientific data, which aim to use the web as a distributed data space by relying on stateless API ecosystems (Heath and Bizer, 2011). That approach, followed here, contrasts with the conventional, less scalable aggregation of raw data in a single backend. By engaging a client-side SDK, computational orchestration of remote data backends can be greatly facilitated by the development of serverless 'Web APIs' (Almeida et al., 2019). This approach's scalability is particularly apparent when engaging distributed data resources in real-time, as illustrated by SDKs developed to track coronavirus mortality data (Almeida et al., 2021).

## Methods and Results

The mSigSDK is a Software Development Kit that targets the operation of mSigPortal APIs. This approach involves abstracting a web portal as an API ecosystem, where the portal's own interactive analytics tools use the same APIs that the mSigSDK operates. The mSigSDK utilizes modern JavaScript ES6 standards (EcmaScript 6 Modules) to fully modularize its components and enable modularized use in any web computing environment without any need for download, installation, or configuration. For instance, multiple developers can seamlessly integrate any number of the visualizations and functionalities provided by mSigSDK into their own web page, using a single import statement. Moreover, adhering to ES6 standards facilitates the extension of mSigSDK to integrate other GWAS APIs by adding new methods or generalizing existing ones in a shared SDK.

The development of the SDK began with an examination of mSigPortal's APIs at https://analysistools.cancer.gov/mutational-signatures/#/apiaccess. The mSigPortal's landing page provides a swagger.io-based API sandbox and data model under 'API Access.' We then created methods that generate visualizations by mapping the data obtained via the APIs to available graphics libraries. All interactions with the backend are mediated through the HTTP APIs provided by mSigPortal.

The mSigSDK depends on several libraries, including the graphics libraries Plotly and AMCharts, a caching library, localforage, which operates the browser's native NoSQL database IndexedDB, and a dimensional reduction library UMAP-js for dimension reduction using Uniform Manifold Approximation (UMAP) method (see Github link under Availability).

The code used to create this SDK is both managed and disseminated to the public domain on Github Pages (i.e. versioned hosting via github.io). Several companion Observable Notebooks were developed to showcase the variety of graphics (Figure 1) that can be rendered from the JSON data structures obtained from mSigPortal. For off-line analysis, there is also an option to download the corresponding data tables used to construct the graphs in CSV format.

## Conclusion

The National Cancer Institute and Division of Cancer Epidemiology and Genetics designed the mutation signature portal with a focus on interoperability, using HTTP REST APIs to orchestrate communication with backend data services. This allowed the creation of the mSigSDK to assist the development of web-based interactive analytics tools that can be invoked without the need for off-line data downloads or software installations, simplifying the logistics and protecting the privacy of the analysis: the data is processed entirely on the client-side (in-browser). As a consequence, mSigSDK can be used from any web-stack environment, from web applications to data portals, eliminating the need for data aggregation on the server-side.

The primary goal of this project was to create a software engineering framework for interoperability that follows FAIR principles to the extent that it can also operate other mutation signature portals and relevant data sources as they become available, as illustrated by the International Cancer Genome Consortium ([icgc.org](icgc.org) ) and NCI's Genomic Data Commons ([gdc.cancer.gov](gdc.cancer.gov)). As mutation signature analytical platforms mature to the point that the respective SDKs interoperate, a more organic route to the emergence of data commons (Grossman et al., 2016) becomes a distinct possibility.

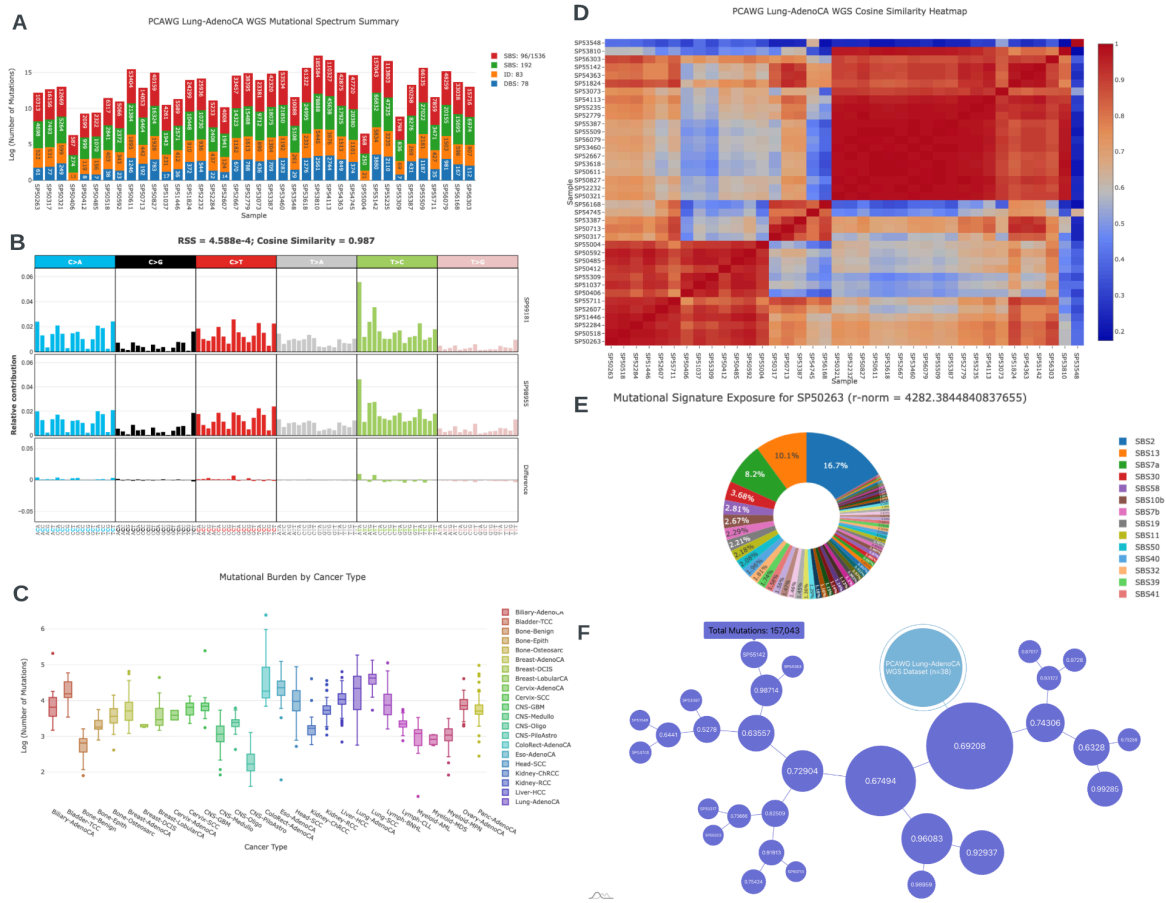

**Figure 1** - Exploring the variety of interactive analytic visualizations generated with mSigSDK. **(A)** Summary data for the PCAWG Lung-AdenoCA, whole genome sequencing dataset. The graph shows the mutational data contained for each of the patients in the dataset **(B)** An example graph comparing the mutational spectra of two samples ("SP99181" and "SP98955") from the PCAWG Lung-AdenoCA, whole genome sequencing dataset. **(C)** A boxplot of the log of the mutational count of each cancer type in the PCAWG dataset. **(D)** The non-negative-least-squares extracted exposures of the PCAWG Lung-AdenoCA, whole genome sequencing dataset, using SBS96 COSMIC v3 Signatures referencing the genome GRCh37. The sample names and signatures of the heatmap have been sorted via double clustering using the unweighted pair group method with arithmetic mean (UPGMA) algorithm. **(E)** A pie chart illustrating the relative exposure values for a particular patient,

SP50263, in the dataset extracted via the non-negative least squares (NNLS) algorithm. The reference signatures used for this example is the SBS96 COSMIC v3 Signatures referencing the genome GRCh37. The r-norm value is the sum of squared residuals for the NNLS decomposition. **(F)** A force directed tree of the patients within the PCAWG Lung-AdenoCA dataset. The tree was generated via the UPGMA algorithm with a cosine similarity distance matrix. The numbers on each node represent the cosine-similarities of the two clusters beneath it. The size of each node is directly proportional to the total number of mutations of the samples under the node.

## Funding

This work was funded by the National Cancer Institute (NCI) Intramural Research Program (DCEG/Episphere).

**Citations**